\documentclass[fleqn,3p,final]{elsarticle}
\usepackage{graphicx} \usepackage{amsmath}
\usepackage{nicefrac}
 \usepackage{xcolor,colordvi}

\usepackage[bitstream-charter]{mathdesign}
\usepackage[T1]{fontenc}

\journal{J. Comput. Phys.}

\usepackage{sidecap}
\sidecaptionvpos{figure}{c}
 \usepackage[
         pdftitle={Parallelized event chain algorithm for dense 
   hard sphere and polymer systems},
          pdfauthor={Tobias Kampmann, Horst-Holger Boltz, Jan Kierfeld},
          colorlinks=true]{hyperref}

\AtBeginDocument{%
\hypersetup{linkcolor=green!60!black,citecolor=gray,%
urlcolor=green}
}
\let\oldeqref\eqref
\renewcommand{\eqref}[1]{\textcolor{green!60!black}{\oldeqref{#1}}}

\let\oldcite\cite
\renewcommand{\cite}[1]{\textcolor{gray}{\oldcite{#1}}}
\usepackage[labelfont=bf]{caption}

\fontsize{9}{12}
\begin{document}

\begin{frontmatter}
\title{Parallelized event chain algorithm for dense 
   hard sphere and polymer systems}

 \author{Tobias A. Kampmann}
 \ead{tobias.kampmann@tu-dortmund.de}
 \author{Horst-Holger Boltz}
 \author{Jan Kierfeld}
 \ead{jan.kierfeld@tu-dortmund.de}
 \address{Physics Department, TU Dortmund University, 
 44221 Dortmund, Germany}

 \date{\today}

 \begin{abstract}
We combine parallelization and cluster Monte Carlo for 
hard sphere systems and present a parallelized event chain 
algorithm for the   hard disk system  in two dimensions.
For parallelization we use a spatial partitioning approach into 
simulation cells. 
We find that it is crucial for correctness 
to ensure detailed balance on the level of Monte
Carlo sweeps by  drawing the starting sphere 
of event chains within each simulation cell with replacement.
We analyze the performance gains for the parallelized event chain
and find a criterion for an optimal degree of parallelization. 
Because of the cluster nature of event chain moves massive 
parallelization will not be optimal. 
Finally, we discuss first applications of the event chain algorithm to 
dense polymer systems, i.e., bundle-forming solutions of 
attractive semiflexible polymers. 
 \end{abstract}
\end{frontmatter}

 \section{Introduction}

Since its first application to a hard disk system \cite{metropolis1953},
Monte Carlo (MC) simulations have been applied to practically all types of 
models in statistical physics, both on-lattice and off-lattice. 
MC samples microstates of a thermodynamic ensemble 
statistically according to their Boltzmann weight. 
It requires knowledge of 
microstate  energies rather than interaction forces. 
In its simplest form, the Metropolis MC simulation  \cite{metropolis1953},
a MC simulation is easily implemented for any system by 
offering a new microstate to the system and accepting or rejecting 
according to the Metropolis rule, which is based on the Boltzmann factor. 
The Metropolis rule guarantees detailed balance.
Typical local MC moves, such as single spin flips in spin systems 
or single particle moves in off-lattice systems of interacting particles, 
are often  motivated by the actual dynamics of the system.
Sampling  with local moves 
can become slow, however,  under certain circumstances, most notably, 
close to a critical point, where large correlated regions 
exist, or in dense systems, where acceptable moves become rare.

There have been two routes for major improvement of MC simulations 
to address the issues of critical slowing down and
dense systems, namely cluster algorithms and parallelization:

(i) One route for improvement is {\em cluster algorithms}, 
  which go  beyond local Metropolis MC. 
  Such methods construct MC moves of large {\em non-local}
 clusters. Ideally, clusters are generated in a way that 
 the MC move of the  cluster is performed {\em rejection-free}. 
 Clusters have to be sufficiently large and 
 cluster building has to be sufficiently fast to gain performance. 
 For lattice spin systems, the  Swendsen-Wang \cite{Swendsen1987}
 and  Wolff \cite{Wolff1989} algorithms  
 are the most notable cluster algorithms with 
 enormous performance gains 
 close to criticality, where they reduce  the 
 dynamical exponent governing the  critical slowing down.

 For off-lattice interacting particle systems, the simplest of which 
 are dense hard spheres,  cluster algorithms have been proposed 
 based on different types of cluster moves. 
 In Ref.\ \citenum{Dress1995}, a cluster algorithm based on pivot 
 moves has been proposed, which was applied to 
 different hard core systems \cite{Buhot1998,Santen2000},
 and variants formulated for soft core systems \cite{Liu2004}.
 In Ref.\ \citenum{krauth2009}, 
 the event chain (EC) algorithm has been proposed, which  
 generates large clusters of particles in the form of a  chain 
 of particles, which are  moved simultaneously and
 rejection-free. ECs become  long in the dense limit,
 which significantly reduces autocorrelation times.

(ii) The other route is  {\em parallelization}, 
 as multi-processor computing has become 
 widely available both in the form of multiple CPUs and, in recent years,
 in the form of graphic processing units (GPUs). On GPUs, 
 ``massively'' parallel algorithms 
 can significantly improve performance of simulations. 
 For massively parallel computation on GPUs,
 the algorithm has to be data-parallel to gain performance, 
 i.e.,  the simulation system 
 has to be dividable into pieces, which can be updated 
 independently accessing a limited shared memory. 
 This  has  been achieved very efficiently for molecular dynamics (MD) 
 simulations  \cite{Anderson2008,VanMeel2008} and 
 MC simulations \cite{Preis2009}. It is an 
 ongoing effort to massively parallelize other simulation algorithms.

It seems attractive to combine both strategies and search for 
parallelization options for cluster algorithms. 
For conventional  Metropolis MC based on single spin/particle moves or 
MD simulations with finite range of interactions,
the parallelization strategy typically consists in spatial partitioning 
of the system into several domains, on which the algorithm 
works independently, i.e., in a data-parallel manner. 
Such algorithms can also be massively parallelized for GPUs.
For cluster algorithms the suitable parallelization strategy is 
less clear. 
If a spatial decomposition strategy is to be used, it 
must be applied to the cluster selection and cluster identification. 
This has been achieved  for 
Swendsen/Wang and Wolff algorithms for lattice spin models 
recently \cite{Komura2012,Komura2014}, and these algorithms have been 
implemented with efficiency gains on GPUs.

The EC algorithm for dense hard sphere system \cite{krauth2009}
relies on a sequential selection of a chain of particles as the cluster 
to be updated (and will be explained in more detail below), 
which makes massive parallelization difficult. 
In this article, 
we want to investigate a strategy to apply spatial partitioning
into independent simulation cells
as parallelization technique to the EC algorithm for 
hard sphere systems in order to combine performance gains 
from cluster algorithm and parallelization.  
A similar approach has been proposed in Ref.\ \citenum{kapfer2013}. 
Here, we also systematically test parallelized EC
algorithms for correctness and efficiency using the well-studied example 
of  two-dimensional hard disks.
As a result, we find  that for the parallel
EC algorithm to work correctly it is crucial 
how the starting points of ECs are selected 
during a sweep  in a simulation cell.

Moreover, the most efficient 
parallelization will not be massive; the 
scalability will be 
limited by  the nature of the EC algorithm itself.
The EC algorithm is most efficient if  EC clusters
have an optimal  size \cite{krauth2009}, which is 
related to the particle density. 
If simulation cells become too small
compared to the typical size of EC clusters, 
parallelization becomes inefficient.
The parallel EC algorithm will, therefore, be best suited for 
 multicore CPUs with shared memory.

We present and systematically 
test the parallel EC  algorithm in detail 
in the context of the hexatic to liquid transition in 
two-dimensional  melting of a system
of hard disks and give an
outlook to dense polymeric  systems in the end.

\subsection{Two-dimensional melting}

Two-dimensional melting, especially in the simplest formulation of a system
of hard (impenetrable) disks, is a fascinating example of a classical phase
transition. Hard disks are an epitome of a system that is easily described
and quickly implemented in a (naive) simulation, but very hard to tackle
analytically. Hard disks have 
been subject of computational studies ever since the
seminal works of Metropolis {\it et al.} \cite{metropolis1953}, which
can be regarded as a starting point for MC simulations and of the
area of computational physics as a whole.

For two-dimensional melting, there has been a long debate on the nature 
of the phase transitions leading from the liquid to the solid phase 
(see, e.g., Ref.\ \citenum{Strandburg1988} for a review). 
In two dimensions genuine long-range positional order is not possible 
because of thermal fluctuations, but a two-dimensional fluid 
with short-range interactions 
can only condense into a solid phase with 
algebraically decaying positional correlations.
The KTHNY-theory \cite{kosterlitz1973,Halperin1978,Young1979} 
describes two-dimensional melting as
defect-mediated two-step melting process:
Starting from the quasi-ordered solid in a first transition 
dislocations unbind, which destroys the translational order 
\cite{kosterlitz1973}
resulting in a so-called hexatic liquid, which remains 
orientationally ordered.
In a subsequent second transition, disclinations unbind destroying also the
remaining orientational  order \cite{Halperin1978,Young1979}.
Both transitions are continuous phase transitions of Kosterlitz-Thouless type. 
Alternatively, a weak first order phase transition has been 
discussed, where a liquid phase with 
lower free energy appears before the 
instability of the ordered phase  with respect to defect-unbinding
sets in. Both phases then coexist in a region in parameter space.  
Simulations on hard disks in two dimensions gave indecisive results
regarding this issue for many years
(see Ref.\ \citenum{Mak2006} for a discussion).

The EC algorithm  helped to settle this issue for 
the two-dimensional hard disk system.
In Ref.\   \citenum{krauth2011} 
it was shown that the transition from the hexatic to the 
liquid phase is a weak first order transition by 
identifying a region of phase coexistence and a characteristic 
pressure loop if the pressure $P$ is measured as a function of the
particle density.
No such loop was detected between the hexatic and solid phase  
with quasi-long-range positional order indicating that 
the hexatic to solid transition is  continuous.
In Ref.\ \citenum{engel2013}, these results were corroborated 
by massively parallel local MC and MD simulations.

\section{Model}

We simulate a gas of $N$ hard disks of diameter $\sigma$ in an $L\times L$ box
with periodic boundary conditions. As velocities trivially decouple from
positions, we restrict ourselves to the latter. 
Hard disks are an athermal system and the particle density $\rho = N/L^2$ or 
the occupied volume fraction $\eta$ defined as
\begin{equation}
 \eta = N \pi \sigma^2 /4L^2 = \pi \sigma^2\rho /4 \text{.}
\end{equation}
is the only control parameter for the phase behavior of the system.  The
system is in a disordered liquid phase for $\eta<\eta_{lh}\approx 0.7$, in a
hexatic phase for $\eta_{lh}<\eta<\eta_{hs}\approx 0.72$ and, finally, in an
ordered phase for $\eta_{hs}<\eta<\eta_{hcp}=\pi/ 2\sqrt{3} \approx 0.9069$,
where the upper bound $\eta_{hcp}$ corresponds to a hexagonal close packaging
of spheres \cite{krauth2011,engel2013}.

We will explore the new parallelized version of the EC
algorithm using the example of the 
liquid to hexatic transition following 
Refs.\ \citenum{engel2013,anderson2013}. 
The hexatic phase is characterized
by  bond-orientational order, which gives rise to a finite
absolute value of the hexatic order parameter $\Psi$ given by
\begin{align}
  \Psi &= N^{-1} \sum_k \Psi_k  \label{eq:Psi} \\
&\text{with } \Psi_k = \sum_{\langle k,\,l\rangle} 
 \frac{\exp\left(6 \text{i} \varphi_{k,l}\right)} 
   {n_k} , \nonumber
\end{align}
where the first summation extends over all particles $k$ and the second one
over all of their $n_k$ next neighbors $l$.  The angle $\varphi_{k,l}$ is the
orientational angle between the vector from the particle $k$ to its neighbor
$l$ and a fixed reference axis (the x-axis in our implementation).

To decide which particle pairs constitute next neighbors a Delaunay
triangulation (the dual graph of the  Voronoi diagram) is used, see e.g.\ 
Ref.\ \citenum{deBerg2000}. In the implementation we made use of the CGAL
library \cite{cgal}. Still the triangulation leads to a high computational cost
for the measurement of $\Psi$ in a simulation, which we therefore avoid as far
as possible. 
However, the hexatic order parameter $\Psi$ is very
useful to quantitatively track the change in the state of the system during
the simulation. For this purpose we define the autocorrelation time $\tau$
as characteristic time scale in the exponential 
  decay of the $\Psi$-autocorrelation
\begin{equation}
C( \Delta t) = \frac{\langle \Psi^*(t) \Psi(t+\Delta t) \rangle}
   {  \langle \Psi^*(t) \Psi(t) \rangle} \sim \exp{(-\Delta t/\tau)} \text{.}
\label{eq:tau}
\end{equation}
Note that the phase of $\Psi$ for a given state depends on the choice of the
arbitrary reference axis and therefore the average $\langle \Psi \rangle$
should be zero, which we already incorporated in the definition of the
autocorrelation. This is a useful check to decide if the measurement was
performed over sufficiently long time or, correspondingly, a large enough
ensemble of systems  (we refer to the ``number of steps in
our MC simulation'' as ``time'' for the sake of conceptual simplicity,
and to  the actual time a simulation runs on a computer as ``wall time'').

We use $\tau$ as a measure for the speed of the simulation 
and the efficiency of the sampling. To
precisely check for the correctness of the simulation
we measure the pressure $P$ as a function of the
occupied volume fraction $\eta$. Close to melting,  the pressure is 
extremely sensitive to problems in the algorithm 
and, therefore, very suitable to test new algorithms. 
The pressure in
this hard sphere system is given by the value of the 
pair correlation at contact distance \cite{metropolis1953}, 
 \begin{align}
   \beta P \sigma^2= \sigma^2 \rho \left( 1 + \frac{\pi}{2} \sigma^2\rho
     \lim\limits_{r \rightarrow \sigma+} g(r) \right) \, ,
\label{eq:pressure}
\end{align}
where $\beta = 1/k_B T$ denotes the inverse temperature, 
$\rho = N/V = 4 \eta /\pi \sigma$ the density,
 and $g(r)$ the pair correlation function.
We measure $g(r)$ using a histogram $n(r_i)$ which counts all pair
distances $d$ with $|d - r_i| \leq \delta_r/2$, 
\begin{align}
g(R_i) &= \frac{\langle n(r_i)\rangle}{ N \rho 2\pi R_i \delta_r} \\
&\text{with } R_i = 
   \frac{2\left(r_{i+1}^3-r_i^3\right)}{3\left(r_{i+1}^2-r_i^2 \right)},
\end{align}
and follow the procedure laid
out in Ref.\ \citenum{engel2013}
 to extrapolate $\lim\limits_{r \rightarrow \sigma+} g(r)$. 
We stress the importance of the binning rule, as we experienced that a
different binning rule (e.g.\ $d-r_i \leq \delta_r$) leads to differing
results for $g(r)$ with errors of the order of $\delta_r$. 
This dependence has not been seen in Refs.\
\citenum{anderson2013} and \citenum{engel2013}, 
where it is stated that 
different binning lead to quantitatively similar results.

\section{Algorithm}

\subsection{Traditional local displacement Monte Carlo}

The local displacement MC algorithm  is the simplest way to
implement a Markov chain MC simulation for hard disks. 
Particles are moved by an isotropically distributed 
random vector $\vec r$ whose length is uniformly distributed
between $ 0$ and $ \ell_\text{max}$. The maximal displacement length
$\ell_\text{max}$ determines the acceptance rate of the simulation.
Optimal performance is usually obtained for an 
acceptance rate around  $20\%-50\%$ 
(for a Gaussian move distribution there
is an exact result of an optimal acceptance rate  $23\%$ \cite{roberts1997}), 
 which means in the case of hard disks that $\ell_\text{max}$ 
is of the order of the mean
free path of the particle. Compared to an MD simulation this
method leads to very large autocorrelation
times \cite{engel2013} and, therefore,
is not very suitable to equilibrate larger systems,
in particular, for higher densities.

\subsection{Event chain algorithm}

The EC algorithm introduced by Krauth {\it et al.}
\cite{krauth2009} has been 
developed to decrease the autocorrelation time of a system of hard
disks or spheres. 
It performs rejection-free displacements of several spheres
in a single MC move. 
The basic idea is to perform a large displacement
$\ell$ in a ``billiard'' fashion, transferring the displacement to the next
disk upon collision. 
First, a  starting particle of the EC
and a random direction are chosen. The chosen
particle is then moved in this direction until it touches another
particle; the displacement $\ell_1$ of the first particle is subtracted from
the initial total displacement length and the remaining 
displacement $\ell-\ell_1$ is
carried over to the hit particle, which is displaced next. 
This procedure continues until no
displacement length is left. The total displacement
length $\ell$ is an adjustable algorithm parameter. 
In principle, $\ell$  should 
be as large as possible; however, 
there is a finite $\ell$ beyond which there
is no substantial gain from simulating longer ECs \cite{krauth2009}.

There are two versions of this algorithm which
differ in the direction of the displacement of the hit disk. 
The hit disk is either displaced in the original direction which
is called \emph{straight event chain}  algorithm or in the direction
reflected with respect to the symmetry axis of the collision which is called
\emph{reflected event chain}  algorithm. The straight EC
variant has been
found to sample more efficiently, as it achieves a larger change in the
system state with the same computational effort thus featuring a smaller
autocorrelation time. 
For collision detection a decomposition into square cells 
with a lattice constant of the order of the
mean free path is used. Each disk is assigned to one square cell 
and collision tests are limited to neighboring squares, 
which leads to a complexity of $O(1)$ for each trial move.

The most performant version of the straight EC algorithm, the so called
$xy$-version or irreversible straight EC algorithm, uses only
displacements along coordinate axes and only in positive direction, thus
simplifying various computations, but also breaking the detailed balance
condition (ergodicity is preserved through periodic boundary conditions).  
Our parallel EC algorithm will rely on a partition of the 
system into simulation cells, such that we have fixed rather than 
periodic boundary conditions for each simulation cell.
In this situation abandoning detailed balance is no longer possible,  
and  we only use ECs that fulfill the detailed balance condition.

For optimal parameters the EC algorithm achieves 
a speed up by a factor of roughly $16$ compared to
local MC \cite{krauth2009}. 
This algorithmic improvement has facilitated the extensive study of
large (up to $N=1024\times 1024$) systems and, thus, the clear identification
of the two-dimensional melting process \cite{krauth2011}.

\begin{figure}[th]
\centering
 \includegraphics[width = 0.7\textwidth]{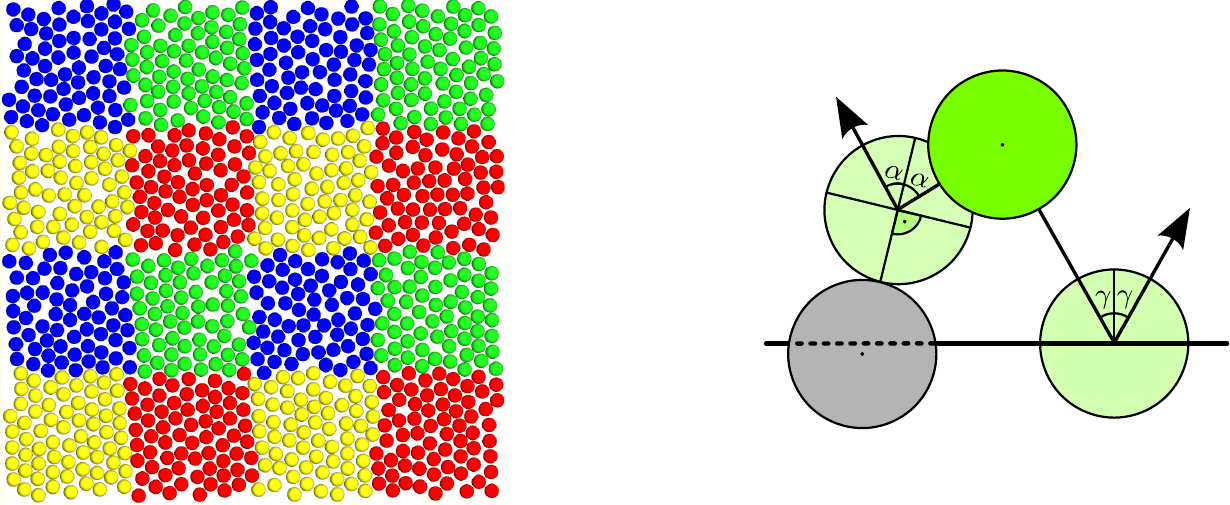}
\caption{
 Left: Scheme of the cell decomposition. {To ensure independence
  of each thread only beads in cells with the same colors 
are moveable at the same time.
  In  every sweep the checkerboard is shifted by a random vector
   to ensure ergodicity} 
 Right: Scheme of {event chain} reflections to handle  cell boundary
   events{, whenever a disk moved
by an EC would move outside the current cell or hit an inactive frozen disk.}
  The angle of reflection of the EC  equals  the angle of incidence.
}
\label{fig:cell_reflection}
\end{figure}

\subsection{Massively parallel MC simulation}

A different approach to overcome the limitations of the local 
MC is to increase the
number of processing units used in the simulation. 
As seen before, the number
of CPUs commonly available in standard workstations would not suffice to
increase the performance to the level reached by the EC
algorithm, even if ideal
scaling is assumed. However,  modern  GPUs can
execute several thousands threads simultaneously and, therefore,
allow for a  massively parallel simulation accessing a 
limited  shared memory.
Such a parallelization comes with its own peculiarities, in particular, 
it has to be taken care that no concurrent memory changes occur.

The  massively parallel MC
algorithm (MPMC) by Engel {\it et al.} 
\cite{engel2013,anderson2013} relies on  the spatial
decomposition into a
checkerboard where four ``cells'' forming a $2\times2$ square
belong to one thread to ensure independent areas for each thread as in Fig.\ 
\ref{fig:cell_reflection} (left). In  every
sweep the checkerboard is {shifted by a random vector}
 to ensure ergodicity and 
a list of all particles in each cell is created and shuffled. 
This step is essential to ensure detailed balance, because in 
every cell a fixed number $n_m$ of particle moves are suggested to 
balance the work load. 
A thread works on  the four cells in a chosen sequential order;
while working on one cell, positions of 
particles in the other cells are frozen.
If a particle attempts to leave a cell
the corresponding move is rejected.

The additional rejection lowers the effectiveness of the sampling, but due to
the massive parallelization the simulation  runs several orders faster than a
simple sequential   local MC algorithm: for optimal parameters a speed up 
by a factor of roughly $148$ was obtained in Ref.\ \citenum{anderson2013} {using
a TESLA K20 GPU with 2496 arithmetic logic units.}

\subsection{Parallel event chain algorithm}

Our goal is to formulate a parallel event chain (PEC) algorithm that uses
commonly available resources (4 to 16 CPUs on a current 
multi-processor machine) which combines aspects from the EC algorithm and
the  massively parallelized local MC algorithm.

For the two-dimensional hard disk system 
our PEC algorithm  for $n$ parallel threads consists of
the following steps in each MC sweep:
\begin{enumerate}
 \item Decompose the system into $4n$ square cells;
   the square lattice is shifted by a random vector. 
 \item Form $n$ blocks of $2\times2$ square 
   cells (cells with different colors in Fig.\ \ref{fig:cell_reflection}).
 \item Randomly shuffle the order of cells in the blocks 
 (select one series of
    colors in Fig.\ \ref{fig:cell_reflection}, 
  e.g.\ blue, yellow, green, red)  
 \item Fork into $n$ threads, each of which acts on one block.
 \item Create a list with $n_m$ EC starting disks
    for each of the four cells in the
    block by drawing {\em with} replacement.
 \item Start one EC at each of the $n_m$ disks in the list.
   While working on one cell, particle positions 
   in the other cells are frozen.
   Synchronize threads when list is done and repeat for all four
   cells in the block.
 \item Go to step 1.\ and generate a new decomposition.  
\end{enumerate}
The generalization to three-dimensional hard sphere systems 
is obvious.

There are two aspects  that turn out to be crucial for the 
correctness and performance of the algorithm: 
(i) how the list of $n_m$ EC starting particles is created
(with or without replacement),
and (ii) how ECs are treated at the cell boundaries
is important for performance. 

\begin{figure}[th]
\centering
\includegraphics{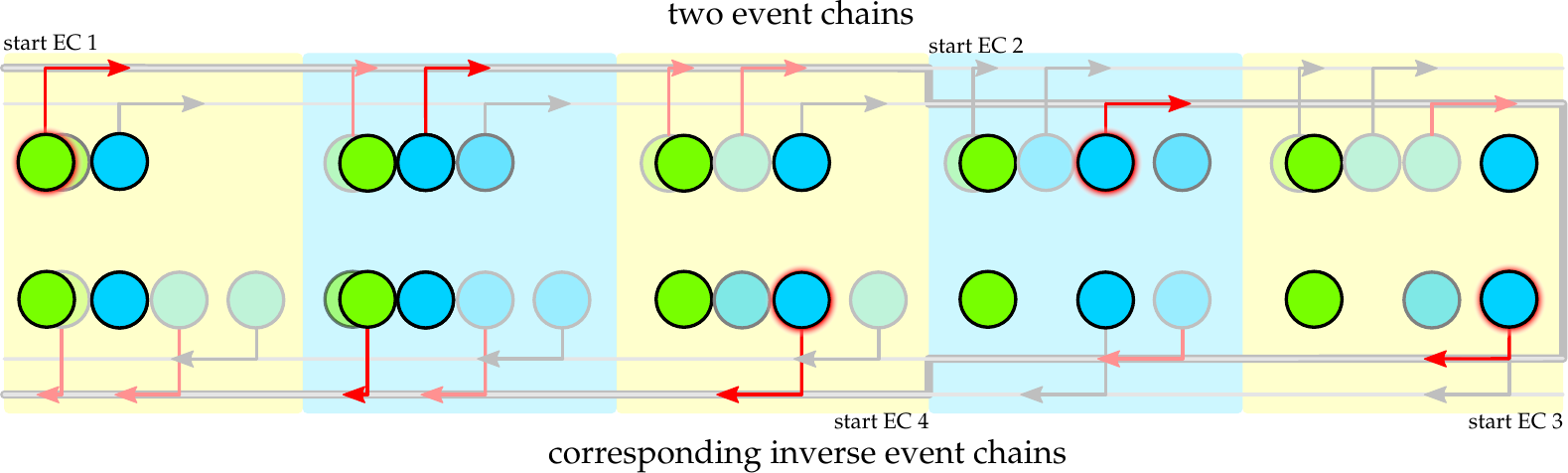}
\caption{ 
 {Influence of event chain starting point distribution on detailed
    balance.  For the sake of simplicity we show a one dimensional system
    containing only two disks.
  A MC sweep then consist of two ECs ($1,\,2$): EC 1 starts at the green 
   disk, EC 2 at the blue disk (upper row). 
     The starting points of event chains are highlighted by a red
    halo. 
 The inverse sweep (lower row) also consists of two ECs  (ECs $3,\,4$).
  Both EC 3 and EC 4 have to start at the blue disk (the end disks 
  of EC 1 and EC 2): 
 Although the starting points of the sweep are
    distinct, for the inverse sweep the event chains need to start on the same
    disk.  To ensure
    detailed balance on the sweep level both sweeps need to be suggested with
    the same probability.
    Therefore a shuffled list with distinct starting points will lead
    to an incorrect sampling, because the inverse sweep cannot be suggested in
    every case.}}
\label{fig:list_shuffling}
\end{figure}

Regarding point (ii), we note that, 
in the actual EC step, the EC must {\em not} leave the
cell, otherwise detailed balance and independence of the parallelly
working threads are not guaranteed. This means that whenever a disk moved
by an EC would move outside the current cell or hit a disk that is
outside the current cell, a separate treatment is required.  In both
cases we think it is most effective to {\em reflect} the EC
 at the cell wall or an  inactive frozen disk at the cell wall
with the angle of incidence equal to the angle of reflection
as shown in Fig.\ \ref{fig:cell_reflection} (right). 
In this way, detailed balance is still
guaranteed and no rejections are introduced. 
In Ref.\ \citenum{kapfer2013}, it has been proposed to {\em reject}
ECs reaching the cell wall or an inactive disk at the cell wall. 
In principle, rejections are also suited to handle these 
situations but will slow down the sampling and also 
impair the load balance among threads.

Regarding point (i), the creation of the list of $n_m$ starting particles 
for ECs is a subtle issue. At first sight, 
it seems favorable to shuffle a list containing $n_m$ {\em distinct}
particles in the cell (as in the MPMC algorithm)
for  a given cell decomposition.
Then the
maximally possible number of distinct EC starting particles is
guaranteed, which seems to promise more efficient sampling. 
However, this use
of a shuffled list of distinct starting particles 
violates detailed balance on the sweep level and gives rise 
to incorrect results (unless $n_m=1$). To see
this we note that the inverse of an EC with displacement $\ell$
starting at disk $a$ with direction $\vec{r}$ and ending at disk 
$b$ is an EC
 with the same displacement $\ell$ and the inverse direction $-\vec{r}$
starting at $b$ and ending at $a$. For a sweep with multiple ECs
starting this means that the sets of starting particles
 of the original and the
inverse sweep need {\em not} to be permutations of each other. 
This is, however,   what
is assumed when using a shuffled list of distinct particles 
(and is valid for local displacement MC).
For detailed balance, all $n_m$ starting particles have to be selected 
with the same equal probability among {\em all} particles in the cell. 
Therefore, one has to create the list of starting particles 
by drawing {\em with} replacement (resulting in starting particles 
which are not necessarily distinct). We give a
simple example for a situation where this becomes relevant in Fig.\
\ref{fig:list_shuffling}. The different statistics -- either drawing with
replacement or without replacement (shuffling) -- of starting particles 
become more relevant the larger the list length $n_m$ is and both  
list types become  identical for $n_m=1$.

\begin{figure*}
\begin{center}
\includegraphics{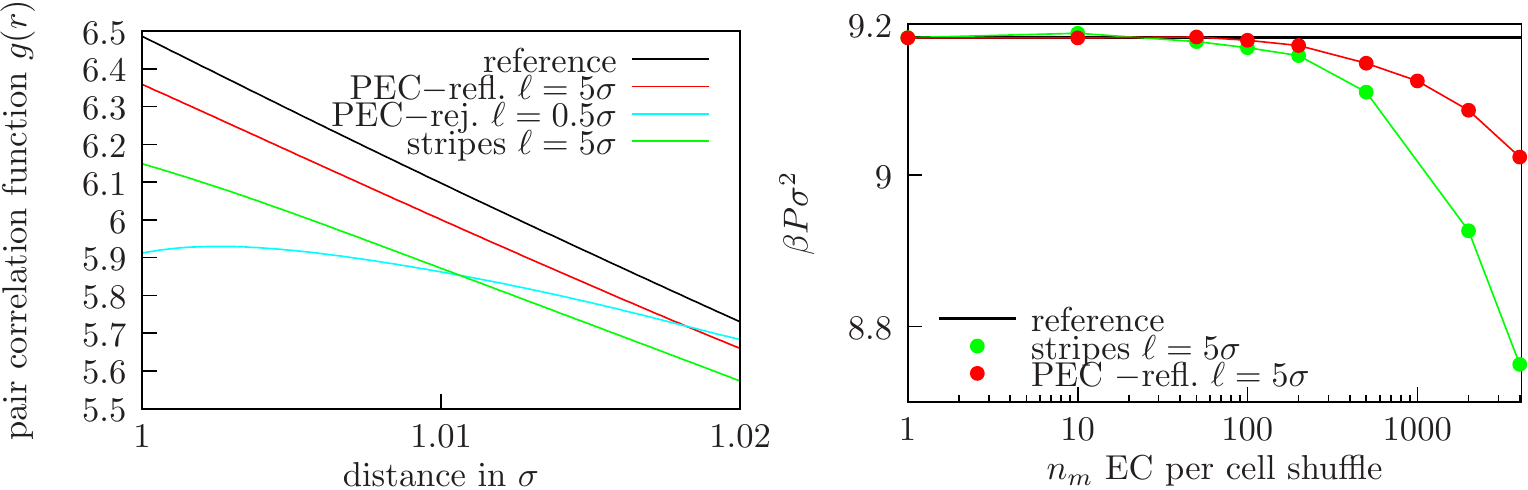}
 \caption{
Left: {Influence of the validity of detailed balance on} the pair correlation
function $g(r)$ for several
{incorrect} simulation variants of the EC  algorithm. 
As a reference,  we show {the correct} $g(r)$ as obtained from a simulation 
with local MC moves (black).
All variants of PEC algorithms using {\em shuffled lists of distinct 
EC starting particles} 
with list  length  $n_m > 1$ give incorrect
results: 
``PEC-refl.''  uses EC reflection, ``PEC-rej.'' uses EC rejection at the
cell boundaries  and a checkerboard decomposition;
``stripes'' uses EC rejection at the boundaries and a striped
decomposition as in Ref.\ \citenum{kapfer2013}. 
On the other hand, {\em all}
corresponding curves for algorithms with 
 starting particle lists drawn with replacement
  lie strictly on top of the reference curve ({and are not shown
  for clarity of the figure}). 
\newline 
Right: Pressure $\beta P \sigma^2$ as a function of the number
$n_m$ of started 
ECs per cell shuffle for different algorithms using
shuffled lists for the starting particles. 
The colors correspond to the colors on the left. 
Only for $n_m=1$, algorithms with shuffled lists become 
correct and the  pressure assumes the reference
 value.
\newline 
All simulations are performed for $N=256^2$ disks at occupied 
volume fraction $\eta = 0.708$.
}
\label{fig:paar_corr}
\end{center}
\end{figure*}

In order to test the importance of the different statistics of
starting particles for different list lengths $n_m$, 
we measure the pair correlation function $g(r)$ and obtain 
the pressure $P$ using eq.\ \eqref{eq:pressure} (for a system containing 
$N=256^2$ disks at occupied volume fraction $\eta = 0.708$). 
Finally, we compare our results with standard 
MC simulations with local displacement moves as a reference. 
We compare PEC algorithms using shuffled lists of distinct 
EC starting particles with PEC algorithms 
using lists generated by drawing particles with replacement;
for both types of algorithms we consider both variants with reflection 
and rejection of ECs at the cell boundaries.

We find that all PEC algorithms starting $n_m>1$ ECs 
at particles drawn {\em without} replacement, i.e., 
by list shuffling give 
incorrect  pair correlation functions $g(r)$.
The  pair correlations  differ from the expected
behavior for small $r\approx \sigma$ both in their functional form,
see Fig.\ \ref{fig:paar_corr} (left), and in the 
resulting value for $P$, see Fig.\ \ref{fig:paar_corr} (right),
which is obtained from the limiting value  of the pair correlation
at  $r\rightarrow \sigma+$ according to eq.\ \eqref{eq:pressure}. 
Differences in the pressure $P$ increase with increasing $n_m$
for as shown  in Fig.\ \ref{fig:paar_corr} (right). 
Only for the trivial list length $n_m=1$, i.e., 
with only EC started in each cell,
there is no difference 
between a list generated by drawing 
with replacement and generated a list
generated by shuffling, and all algorithms converge to the 
reference result for $P$, see Fig.\ \ref{fig:paar_corr} (right).

For all list lengths $n_m>1$, 
all PEC variants  using starting particle lists generated {\em with} 
replacement  yield the correct
reference result for $g(r)$  from the local MC simulation.

We also  examined
the importance of  replacement in the starting particle list
for a parallelization of the EC algorithm through a decomposition
into striped cells as proposed in Ref.\ \cite{kapfer2013}.
Similar to the PEC with 
checkerboard cell decomposition we find 
that only drawing starting particles with replacement
yields the correct pressure as also shown in Fig.\ \ref{fig:paar_corr}.

The dependence of the pressure and pair correlation function on algorithmic
details (choice of $\ell$, reflection/rejection, squares/stripes)
for the incorrect algorithms using  shuffled
lists stems from the different ending point statistics of the ECs and
is not easily explained on a quantitative level.

\begin{SCfigure}
\includegraphics{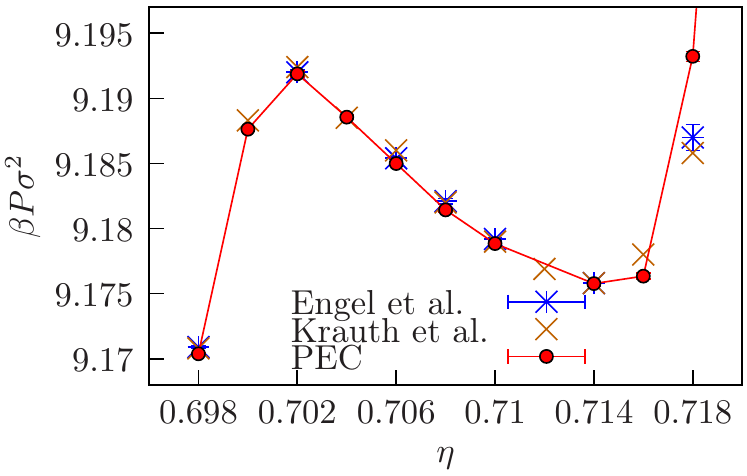}
\caption{
Phase diagram for the liquid-hexatic transition of hard spheres. We
compare the results of the three algorithms MPMC (blue, data taken from 
Ref.\ \citenum{engel2013}), 
sequential EC (brown, data taken from 
Ref.\ \citenum{engel2013}) and PEC (red)
for a system with  $N=256^2$ particles. 
Within the estimated simulation errors
the algorithms yield the identical pressure density dependence.
At large $\eta$, the PEC accuracy is limited by smaller running times. 
We run 30 independent simulations to estimate errors.
}  
\label{fig:phase_diagram}
\end{SCfigure}

For a more extensive quantitative check of the correctness 
of our algorithm, we simulate a system with $N=256^2$ particles 
in the range of occupied volume fractions
$\eta = 0.698\ldots 0.718$, which is the regime of coexisting densities 
at the transition between hexatic and liquid phase as reported in 
Refs.\ \citenum{krauth2011} and \citenum{engel2013}. We 
compare the measured pressure with literature values from these 
references in Fig.\ \ref{fig:phase_diagram}. 
We find good agreement with deviations only at the highest 
packing fractions $\eta$, which is not a problem of our PEC algorithm 
itself: For large $\eta$
 autocorrelation times become very large;
simulations in Refs.\ \citenum{krauth2011} and \citenum{engel2013} 
had significantly longer running times and, thus, give 
 more accurate values.

{
For sequential event chain implementations of hard disk systems there is a
more effective way to calculate the pressure than by means of the 
pair correlation function and eq.\
\eqref{eq:pressure}, which uses a relation between
   the pressure and  collision
probabilities \cite{michel2013}. We chose to compute the pressure
using the pair correlation function as this puts local MC and event chain
data on the same established footing and does not imply greater
computational cost because we use the pair correlation 
as a source of additional information on the correctness of the sampling,
especially with respect to detailed balance (see Fig. \ref{fig:paar_corr}).
 We suggest that the pressure calculation of Ref.\
\citenum{michel2013}  could be extended 
to PEC sampling by treating reflected event chains using a
method of images and plan to explore this in future work.
}

\section{Efficiency of the parallel event chain}

To optimize the  efficiency of the PEC algorithm, we can 
adjust  the degree of parallelization  $n$, 
the number $n_m$ of ECs started within each cell and the 
length of ECs via the total displacement length $\ell$ 
in the parallelization scheme. 
Parallelization of the EC algorithm will 
only offer an advantage over the sequential EC
algorithm if the autocorrelation time $\tau$ in units of wall time
decreases significantly with the number $n$ of threads.

First we discuss efficiency as a function of 
the number $n_m$ of ECs started within each cell. 
For the applications we have in mind (see below), it is favorable to use an
isotropic distribution of displacement directions, which basically eliminates
the striped geometry proposed in Ref.\ \cite{kapfer2013}
because displacements transverse to the stripe direction
will either be rejected or reflected with very little net displacement. To
sample efficiently there should be at least one displacement per
particle which suggests 
\begin{equation}
   n_m  \sim   N/4n n_{\text{EC}}\sim N\lambda_0/4n\ell,
\label{eq:nm}
\end{equation} 
where $n_{\text{EC}}$ denotes the average number of disks per EC. 
On average, this number is given by  the ratio of the total displacement
length $\ell$  of the EC and the mean free path of disks
$\lambda_0$ (approximately given by the free volume per particle 
$\lambda_0 \approx \sigma (\eta_{hcp} - \eta)/2\eta_{hcp}$ 
in the dense limit close to the close-packing density $\eta_{hcp}$), 
\begin{equation}
   n_{\text{EC}} \sim \ell/\lambda_0.
\label{eq:nEC}
\end{equation}

\begin{figure}
\begin{center}
\includegraphics{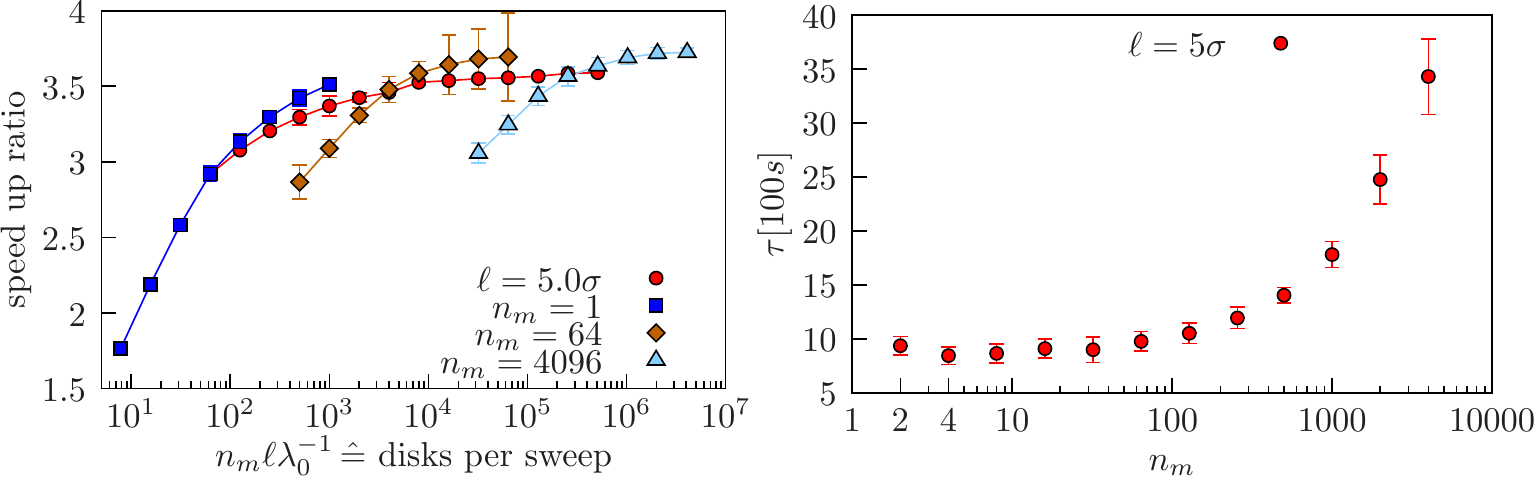}  
\caption{Left: Ratio of the number of moves per seconds (wall time) of the
  parallel and sequential EC algorithm as a function of the total displacement
  length $\ell$ and the number $n_m$ of ECs started per cell configuration as
  a function of $n_m \ell /\lambda_0$ (for $N=256^2$ disks at $\eta
  =0.71$ using $n=4$ blocks of $2\times2$ square cells).  
   The speed {up}
  saturates to the maximal factor $n=4$ both for large $n_m$ and for large
  $\ell$.  Since the required computation time for the synchronization
  overhead rises with $n_m$, it is more favorable to increase $\ell$ rather
  than $n_m$.  The speed up should not depend on other simulation parameters
  such as $\eta$ or $N$.\newline 
  Right: Autocorrelation time $\tau$ of the parallel EC algorithm measured
  in wall time as a function of the number $n_m$ 
   of started ECs per cell (for $N=64^2$ disks at $\eta
  =0.704$ using $n=4$ blocks of $2\times2$ square cells). 
   At $n_m \approx 5$, each disk in a cell is moved once per sweep on average
  according to eq.\ \eqref{eq:nm}. 
   If $n_m$ increases the efficiency of the algorithm decreases
  strongly.}
\label{fig:move_per_sec}
\end{center}
\end{figure}

 For  small values of $n_m$, the relative fraction of
wall time spent on overhead (forking / synchronizing / list creation)
increases.
Reducing the relative overhead by increasing $n_m$  leads to the 
saturation of speed up (in terms of the number of MC moves per wall time)
by parallelization to a maximal value close to $n$ 
as shown in Fig.\ \ref{fig:move_per_sec} (left).
For values of $n_m$ much larger than the value  (\ref{eq:nm}) 
 finite size effects will increase as each decomposition into $4n$ 
cells persists for a long time
(in the limit $n_m\rightarrow \infty$ the system consists of $4n$ 
entirely independent smaller systems). 
Moreover, sampling  $4n$ small cells independently  for a long time 
 is inefficient in removing large  scale correlations extending 
over several cells.
Therefore,  the autocorrelation time $\tau$ of the PEC 
algorithm (measured in wall time)
increases  for large $n_m$ as shown 
 in Fig.\ \ref{fig:move_per_sec} (right).
The optimal choice of $n_m$ according to the results in 
 Fig.\ \ref{fig:move_per_sec} (right)
 agrees with  the criterion \eqref{eq:nm}.

Now we discuss efficiency in terms of 
 the total displacement length $\ell$ and the 
degree of parallelization $n$. 
The efficiency of the  EC algorithm  in general depends crucially 
on the average number of disks per EC  
$n_{\text{EC}} \sim \ell/\lambda_0$.
In Ref.\ \citenum{krauth2009}, it has been found that 
the  straight  EC algorithm has optimal
efficiency for  $\ell/\lambda_0  \sim 10^1...10^3$ (for a system 
size $L\approx 34\sigma $). 
For smaller $\ell$, the efficiency decreases and 
approaches traditional local MC (corresponding to $\ell<\lambda_0$).
For very large $\ell$, ECs comprise large parts of the system
and give rise to motion similar to a collective translation of disks,
which is also inefficient.

\begin{figure}
\begin{center}
\includegraphics{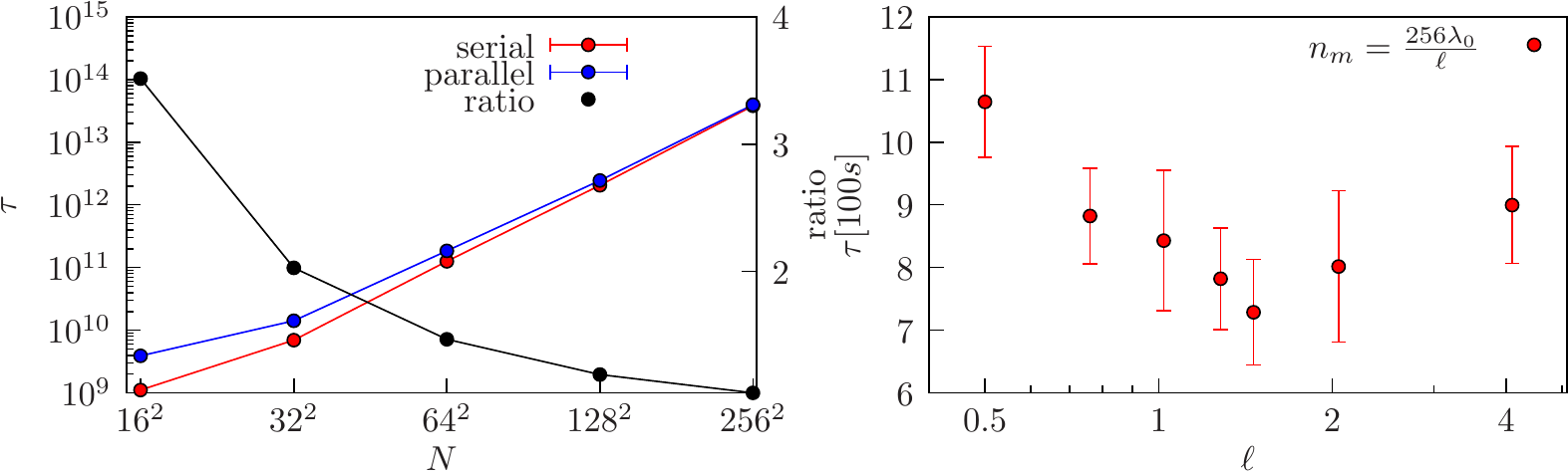} 
\caption{ Left: Autocorrelation time {of the order parameter $\Psi$ for}
parallel and sequential EC algorithm 
(in units of effective local MC moves) as
  a function of the number of particles $N$ for fixed volume fraction $\eta
  =0.71$, total displacement length $\ell = 5 \sigma$ and one EC per cell
    configuration $n_m=1$.  One collision in an EC is treated
    20 effective local MC moves in a local displacement MC
  simulation \cite{krauth2009}.  For larger systems $N \geq 128^2$, the
  autocorrelation time differs only slightly between parallel and sequential
  algorithm. Therefore a good measurement for the efficiency of the algorithm
  is the mere increase in events per wall time shown in Fig.\
  \ref{fig:move_per_sec} (left).\newline 
  Right: Autocorrelation time of the parallel EC algorithm (in units
  of wall time) as a function  of the total displacement length $\ell$ and with
  constant number of moved disks per sweep $n_m n_{\text{EC}}{=} {\rm const}$
   for $N = 64^2$ and $\eta=0.704$. 
  The autocorrelation time exhibits a  minimum for $\ell
  \gtrsim 10 \lambda_0 \approx 0.8$ and $\ell \approx
  {L\lambda_0}/{\sqrt{n}\sigma} \approx 1.28$ in agreement with
   eq.\ \eqref{eq:criterion}.
 }
\label{fig:auto_corr}
\end{center}
\end{figure}

For the efficiency of the PEC algorithm, 
two additional competing effects are relevant to determine
 the optimal values for $\ell$ and the degree of parallelization $n$. 
On the one hand, it is obvious that the reflections at the cell walls 
in the PEC algorithm will impair the sampling of
 configuration space. 
Therefore, the autocorrelation time of the PEC algorithm can be 
significantly increased for small cell sizes, which is equivalent to 
small system sizes for a fixed number $4n$ of simulation cells,
or for long ECs, i.e., large  total displacement
lengths  $\ell$ of the ECs.
This leads to a decrease in efficiency if the EC
length $L_{\rm EC} \sim \sigma n_{\rm EC} \sim \sigma \ell/\lambda_0$ is 
increased beyond the cell size $L/2\sqrt{n}$, see Fig.\ \ref{fig:auto_corr}.
On the other hand, the relative
overhead in computation time from parallelization (due to 
forking/joining/synchronizing threads) is larger for shorter chains.
This leads to a decrease in efficiency for decreasing $\ell$, 
see Fig.\ \ref{fig:move_per_sec} (left).

We conclude that $\ell$ should be chosen within the  window 
$\ell  \sim 10^1...10^3 \lambda_0$ of  optimal values for 
the straight EC algorithm in general, on the one hand, and 
as large as the cell size permits, i.e., according to  
$L_{\rm EC}(\ell)\sim  L/2\sqrt{n}$ or  
\begin{equation}
 \ell \sim \frac{L\lambda_0}{\sqrt{n}\sigma} \sim 
  \lambda_0 \left(\frac{N}{\eta n}\right)^{1/2},
\label{eq:criterion}
\end{equation}
on the other hand. 
If the number $n$ of parallel threads is chosen too large, this 
choice for $\ell$ will drop below the optimal window 
$\ell < 10 \lambda_0$ 
for straight  EC algorithms and efficiency goes down.
Therefore, we propose to adjust the degree of  parallelization
of the PEC algorithm
according to the criterion \eqref{eq:criterion} 
rather than massively parallelize.

In Fig.\ \ref{fig:auto_corr}, we  quantitatively 
investigate the optimization 
of efficiency by adjusting  $\ell$  or the  EC
length $L_{\rm EC} \sim \sigma \ell/\lambda_0$  according to the 
cell size for a degree of parallelization $n=4$. 
In order to  investigate how much the simulation 
can be accelerated by parallelization with $n=4$, 
we measured the ratio of the
autocorrelation time $\tau$ 
of sequential and parallel EC  algorithm as a function 
of  the number $N$ of disks for 
 $\ell= 5\sigma$ and a fixed particle density $\eta=0.71$,
see Fig.\ \ref{fig:auto_corr} (left).
The number $N$ of disks is related to the 
system size by $L= \sigma(N\pi/4\eta)^{1/2}\propto N^{1/2}$.
Because we also work with a fixed number $n=4$ of threads,
$N^{1/2}$ is proportional to the system size $L$ as well as the 
cell size $L/4$.  
For $\eta=0.71$, the mean free path is $\lambda_0 \approx 0.08\sigma$
\cite{krauth2009}, such that $\ell= 5\sigma$ corresponds to 
$\ell/\lambda_0 \approx 62.5$, which is well within the 
window   $\ell/\lambda_0  \sim 10^1...10^3$ 
of optimal  total displacement lengths for EC algorithms. 
In Fig.\ \ref{fig:auto_corr} (left), 
we show the autocorrelation time $\tau$ of the parallel and 
standard sequential EC algorithm in units of MC
moves, where 20 MC moves correspond to one collision in the EC
(using the same convention as in  Ref.\ \citenum{krauth2009}). 
We observe that for smaller systems $N < 128^2$ the
autocorrelation time of the parallel EC  algorithm  exceeds
the autocorrelation time of the sequential EC algorithm.
First, this means that for smaller systems $N <128^2$ corresponding 
to cell sizes $< 34\sigma$ the parallelized EC 
algorithm becomes less efficient. Because for 
$\ell= 5\sigma$ and $\lambda_0 \approx 0.08\sigma$, the 
typical EC length is $L_{\rm EC} \sim \sigma \ell/\lambda_0\sim
62.5\sigma$, this marks also the range of 
cell sizes, which are  comparable to or smaller than 
EC lengths. 
This supports our argument that 
EC lengths $L_{\rm EC}$ should be smaller than 
cell sizes for efficiency of the PEC algorithm.

The results in Fig.\ \ref{fig:auto_corr} (right)
show explicitly that there exists an optimal $\ell$ for the 
PEC algorithm that minimizes the autocorrelation time $\tau$
for a system with  $N=64^2$ disks. 
This  minimum is attained for  $\ell
  \gtrsim 10 \lambda_0 \approx 0.8$ and $\ell \approx
  {L\lambda_0}/{\sqrt{n}\sigma} \approx 1.28$ in agreement with
our above arguments and 
   criterion \eqref{eq:criterion}.

Secondly, the results in 
Fig.\ \ref{fig:auto_corr} (left) show 
that for larger systems $N \geq 128^2$ 
the  mere increase in MC moves per wall time is already a good
measure for the efficiency of the PEC algorithm. 
Therefore, the ratio of the number of MC moves  of the parallel and
sequential EC algorithm,
as shown in  Fig.\  \ref{fig:move_per_sec} (left) 
for a system of $N=256^2$ disks, is also a measure of efficiency. 
In Fig.\  \ref{fig:move_per_sec} (left),  we show 
this ratio as a function of $\ell/\sigma$ and $n_m$.
We find that the speed up ratio is an increasing function 
of both $\ell$ and $n_m$.
For large values of $\ell/\sigma$ or  $n_m$, 
the speed up ratio approaches $\approx 3.9$ from below, which is 
remarkably close to the optimal speed up ratio $4$  
for  $n=4$ threads. 
We conclude that 
the PEC algorithm realizes its efficiency gain 
 by an approximately 
equal autocorrelation time as the sequential EC 
algorithm in units of 
MC moves but  a significant increase in the speed up ratio, 
i.e., the number of MC moves per wall time with the degree 
of parallelization $n$. 

In conclusion, we propose the following  strategy 
of choosing the degree of parallelization $n$, 
the total displacement  $\ell$ and  the number  $n_m$ of ECs started within
each cell to optimize efficiency:
1) Choose a degree of parallelization $n$ such that
$N/\eta n  > 100$ such that $\ell > 10 \lambda_0$
according to the criterion \eqref{eq:criterion}.
We used $n=4$ in our simulations.
2) Choose $\ell=\ell(n)$ according to the criterion  \eqref{eq:criterion}
 such that $L_{\rm EC}$ is comparable to the cell size.
3) Choose the number $n_m=n_m(n,\ell)$ according to \eqref{eq:nm}.

 \section{Application to polymeric systems}

In many applications other than pure hard core systems, 
we have to deal with systems containing
hard core repulsion alongside with other interaction 
in a canonical ensemble.
In Ref.\ \citenum{michel2013}, rejection-free extensions to the
EC algorithm for continuous potentials have been presented
using lifting MC moves and the concept of infinitesimal MC moves. 
We will use a simpler approach, where we 
deal with the hard core repulsion using ECs
and  take into account the other interactions 
using the standard Metropolis algorithm. As
this involves rejection and, thus, a slower sampling of the 
hard core degrees of
freedom the need for  parallelization is even larger than in a gas of hard
disks/spheres. We will show that EC type
MC  moves not only improve the sampling of dense systems 
but can also give rise to a  physically more correct
``dynamics''  in an equilibrium MC simulation of dense 
systems.

We will apply this algorithm to a system consisting of many 
semiflexible polymers, such as actin filaments, which 
are interacting via a short-range attractive potential  
in a flat simulation box in three dimensions.
Under the influence of this attraction,  semiflexible polymers 
tend to form densely packed bundles. 
Actin filament bundles consisting of many semiflexible actin filaments,
which are held together by crosslinking proteins 
are a realization of such a system and represent 
important cellular structures \cite{Bartles2000}. 
In vitro, F-actin bundles 
assemble both in the presence of crosslinking proteins and by 
multivalent counter ions. 
Theoretical work on crosslinker-mediated bundling 
of semiflexible polymers
\cite{Kierfeld2003,Kierfeld2005a,Kierfeld2005,kampmann2013} 
and counterion-mediated binding 
of semiflexible polyelectrolytes
\cite{Borukhov2001} 
show a discontinuous bundling transition
above a threshold concentration of crosslinkers or counterions. 
Related bundling transitions have been found
 for crosslinker-mediated bundling \cite{Benetatos2003}, 
 for counterion-mediated bundling \cite{Shklovskii1999} and 
 for polyelectrolyte-mediated bundling
\cite{DeRouchey2005}.

\begin{figure}
\begin{center}
\includegraphics[width = 0.975\textwidth]{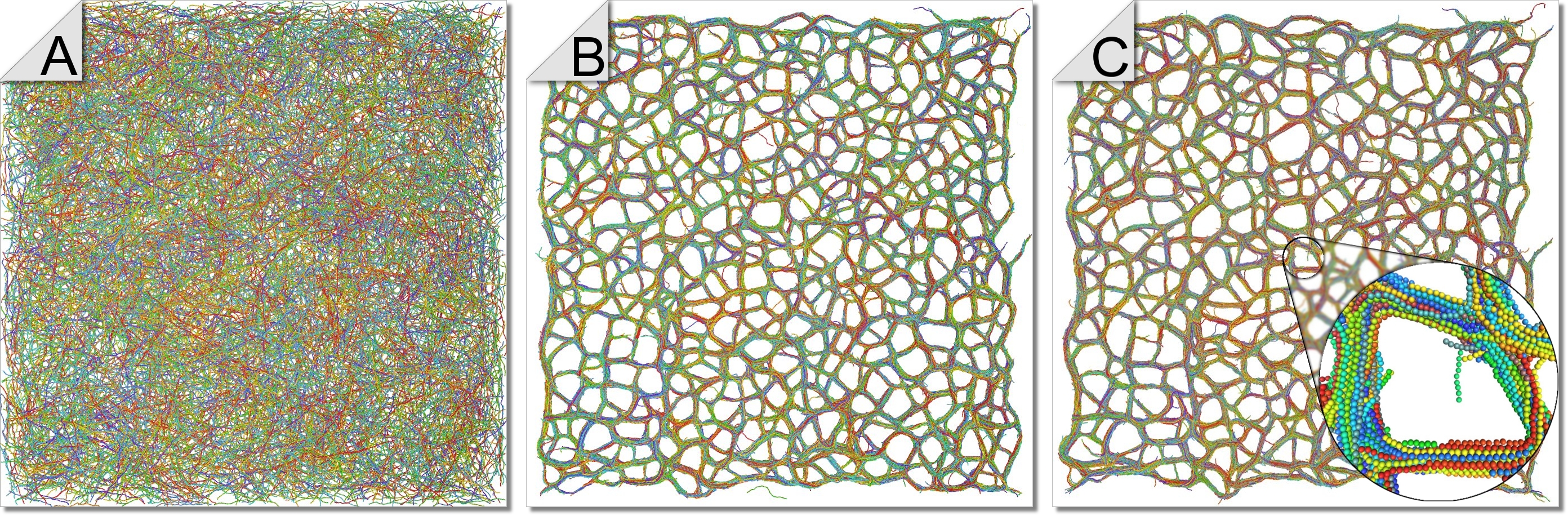}
\caption{
{
  Three  snapshots of polymer system with $10000$ SHCs containing each 
    $50$ beads in a
  flat box of edge lengths $300 \sigma \times 300 \sigma \times 5 \sigma$ 
  with fixed boundaries. Chain colors are arbitrary and  only to 
   distinguish different chains.
 The initial state A of the simulation 
  is prepared with homogeneously and  isotropically distributed chains
  resembling a polymer solution state.
  Snapshot B  is  taken after $3.3 \cdot 10^{6}$ sweeps, snapshot C
  after  $5.1 \cdot 10^{6}$ sweeps (in each MC sweep one event chain 
  is started on each particle).
  Apparently, a stationary state is reached  in 
  snapshot B. 
 }
}
\label{fig:network_snapshot}
\end{center}
\end{figure}

{It is an open question, both experimentally and theoretically, 
what  the actual  thermodynamic equilibrium state  of the bundled 
system is and whether this state is kinetically accessible 
starting from a certain initial condition, for example, 
an unbundled polymer solution.  
Energetically, it is favorable to form a thick single bundle;
entropically, a network of bundles (as shown in 
  Figs.\ \ref{fig:network_snapshot} B and C) might be advantageous.
In both cases,} 
the resulting
bundles of semiflexible polymers are typically rather densely packed.
This causes difficulties in equilibrating bundled structures 
in traditional MC simulations employing local moves. 
In Ref.\ \citenum{Kierfeld2005} MC simulations 
showed evidence for kinetically arrested 
 states with segregated sub-bundles, 
in Ref.\ \citenum{Stevens1999} kinetically arrested bundle networks 
have been observed. Further numerical progress 
requires a faster equilibration of dense bundle structures. 
Ideally, the simulation reproduces the physically realistic 
center of mass diffusion kinetics of whole bundles. 
Networks of bundles have also been observed experimentally in 
vitro for actin bundles with crosslinkers \cite{Pelletier2003} 
and, more recently, in actin solutions where the counterion 
concentration has  been increased in a controlled 
manner \cite{huber2012,deshpande2012}.

Here we report a first  application of 
the EC algorithm to a polymeric system
of semiflexible harmonic chains (SHC)
{
 which are modeled as chains 
of  hard spheres or beads of diameter $\sigma$
 connected by extensible springs with rest length $b_0=\sigma$ and 
spring constant $k$ with an additional 
 three bead bending energy characterized by 
 a bending rigidity  $\kappa$ \cite{Kierfeld2004}.
Beads in different chains also interact with a short-ranged 
attractive square well potential of strength $g$ and range $d$. 
We choose the potential strength $g$ well above the critical value 
for bundling \cite{Kierfeld2003,Kierfeld2005,kampmann2013};
the range  of the attractive square well potential is $d=0.5\sigma$.
}

{
The energy of this system can be expressed as 
\begin{align}
H [{t_i}] = 
  \sum_n \left[ \kappa \sum_{i=0}^{N-2} 
  \left(1-\frac{\vec t_{n,i} \,\vec t_{n,i+1}}{\lvert\vec t_{n,i}\rvert 
      \lvert  \vec t_{n,i+1} \rvert} \right) +
 \frac{k}{2} \sum_{i=0}^{N-1} \left(\frac{|\vec t_{n,i}|}{b_0} - 1\right)^2
  - \sum_{\substack{\langle \{n,i\},\\ \{m,j\}\rangle}} 
     V_{n,m}(\lvert \vec r_{n,i}- \vec r_{m,j} \rvert) \right]\, ,
\end{align}
where $\vec t_{n,i} = \vec r_{n,i+1} - \vec r_{n,i}$ denotes the tangent vector 
between the $i$th and $(i+1)$th bead of the $n$th chain and $V_{n,m}(\lvert
\vec r_{n,i}- \vec r_{m,j} \rvert)$ describes the pairwise attractive
square-well potential and the hard sphere condition.
We present simulation results for $k=100k_BT$,  $\kappa = 20k_BT$
and $g=1k_BT$. 
}

We consider an ensemble of many SHC in a flat box geometry
of edge lengths $300 \sigma \times 300 \sigma \times 5 \sigma$;
this geometry is similar to what is used in microfluidic 
experimental setups in Ref.\ \citenum{deshpande2012}.
For the chosen potential strength, the 
SHCs  bind together and form a locally dense bundle 
resembling a dense hard sphere liquid.
For systems containing {\it many} SHCs
{prepared in an initial state resembling a polymer solution
(see  Fig.\ \ref{fig:network_snapshot} A)}, 
 we obtain network  bundle 
structures, which are very similar to those seen  in experiments for
F-actin \cite{huber2012,deshpande2012}, see 
Fig.\ \ref{fig:network_snapshot} B and C.  
A detailed quantitative investigation 
of these structures will follow in future publications. 
{The network bundle structure is stationary or kinetically 
arrested on the accessible simulation time scales and characteristic 
for the initial state of a homogeneous and isotropic polymer solution
as it is also used in the experiments \cite{huber2012,deshpande2012}.}

To adapt the EC algorithm to the polymeric SHC system,
the hard sphere repulsion is treated by  an
EC. The  additional energies, such as  bending, stretching and
short-range inter-polymer attraction  are treated 
conventionally employing Metropolis MC steps with 
 rejections. First the
entire EC is calculated as if we deal with 
 a hard sphere system. After that
the energy difference between the initial and final
 configuration determines if
the whole EC move is accepted or rejected. 
This leads to a total displacement $\ell$ which determines
the rejection rate. 
By adjusting the total displacement to an optimal value $\ell_{\rm opt}$, 
we can realize  optimal 
acceptance rates around $50\%$.

\begin{figure}
\begin{center}
\includegraphics{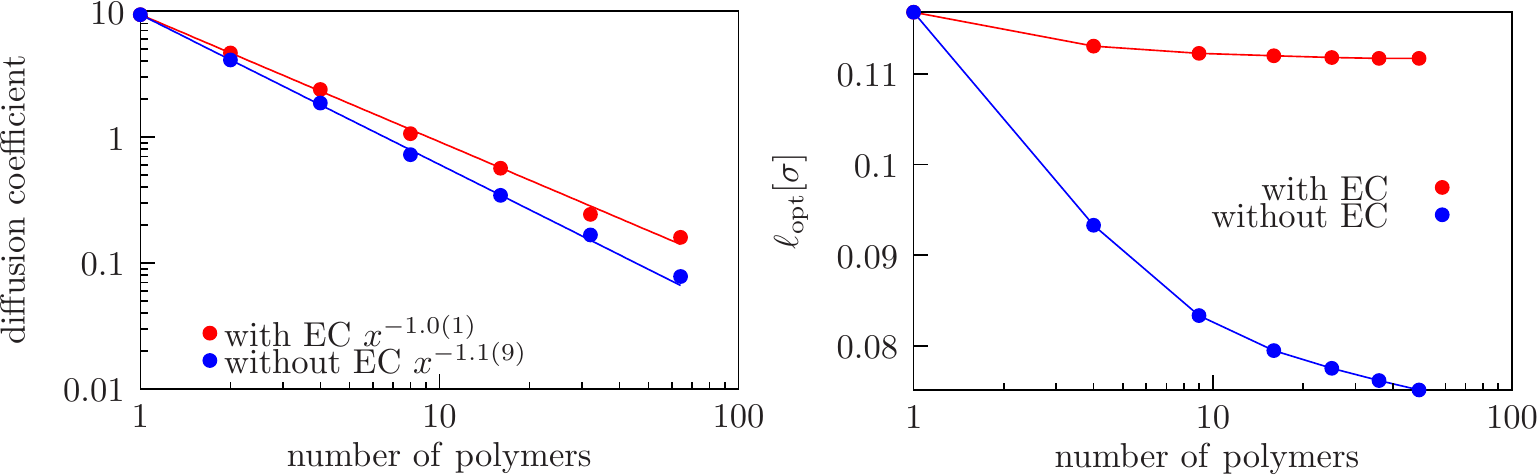} 
\caption{ Left: Diffusion constant of the center of mass of a polymer bundle
  consisting of $N_p$ polymers. {Using an EC algorithm,
   the result $D\sim N_p^{-1}$ characteristic for Rouse
    dynamics is reproduced, which 
   allows for an identification of ``Monte Carlo
    time'' as physical time.} 
   Right: ``Optimal move'' length
  $\ell_\text{opt}$ determined by adjusting the rejection rate to $0.5$ as a
  function of the number $N_p$ of polymers in a bundle. 
{The acceptance
    probability is not reduced by the number of polymers if the event
    chain algorithm is used. This leads to a more effective sampling.}  
}
\label{fig:diff_SHC}
\end{center}
\end{figure}

Our first simulation results  show that such a dense polymeric 
system can also benefit from the EC
algorithm. First, we measure the diffusion coefficient $D$ for a bundle of
polymers depending on the number of polymers $N_p$ while interpreting the
number of MC sweeps as time. 
{
The overdamped time 
evolution of the bundle (neglecting hydrodynamic effects) is described 
by Rouse dynamics, which gives 
a diffusion coefficient 
 decreasing  as $D\sim N_p^{-1}$ \cite{DoiEdwards}. 
}
Applying single displacement local MC moves  to such a system leads to
an unphysically slow  diffusion behavior with 
$D\sim N_p^{-1.2}$,  while the EC algorithm
leads to the correct behavior of the  diffusion constant as shown in 
Fig.\ \ref{fig:diff_SHC}. 
{This implies that the event chain as a cluster
move not only allows for a more efficient sampling in this system, but also
incorporates some essential features of the true
dynamics, in that it allows (at least on a coarse scale) for an identification
of ``Monte Carlo time'' (number of performed sweeps) and physical time. For
the extraction of quantitative information a gauge would be needed, e.g. an
otherwise measured (experiment, MD simulation) diffusion constant.}
The range $d$ of the attractive
square well potential is $d=0.5\sigma$ and, therefore,  rather large
compared to the mean free path of densely packed disks
examined before. For smaller potential ranges $d$ we expect 
the bundle diffusion  to exhibit a stronger  slow down  without 
EC moves.

Furthermore we determine how the optimal displacement 
 length $\ell_\text{opt}$ depends on the number $N_p$ of
polymers in  a bundle. 
For an efficient simulation, $\ell_\text{opt}$ should 
{\it  not} depend on the thickness of a
bundle. Figure \ref{fig:diff_SHC} shows 
that $\ell_\text{opt}$ is drastically decreasing with $N_p$ for
local single displacement MC moves, whereas  it is almost  independent of
$N_p$ for the EC algorithm.
 For smaller potential ranges $d$ we expect 
  a faster decrease in  $\ell_\text{opt}$. Again, we expect the 
  decrease to  be much stronger 
   without EC moves.

These two results can be interpreted such 
that local single  displacement moves do not produce a realistic 
{Rouse} dynamics including center of mass diffusion of whole bundles and do 
not allow to interpret the MC sweep number as a realistic time. 
Furthermore,  first
simulations suggest that even if the EC length is not much longer
than a single displacement the speed up in equilibration time can be very
high. 

With the   parallelized version of the  EC algorithm 
it is possible to 
{reach a stationary quasi-equilibrium state for} large
system sizes as shown in Fig.\  \ref{fig:network_snapshot},
which allows us  to compare with experimental systems.
The snapshot exhibits the same bundle network features 
as observed in experiments  \cite{huber2012,deshpande2012}, 
namely the formation of a network of bundles as the final state 
of the aggregation process. The network exhibits a polygonal 
cell structure, which is also observed in Refs.\
\citenum{huber2012,deshpande2012}. Further quantitative comparison 
between simulation and experiments will be performed in 
future investigations. 
{The efficient sampling technique using the PEC 
 will be useful to adress the question 
whether the experimentally observed network states are 
true equilibrium states 
(due to their higher entropy in comparison to the energetically
 favorable single bundle) 
or  are kinetically trapped metastable states
and how the formation of a network depends on the global 
parameters (such as density and contour lengths of the  polymer chains) 
and initial conditions.
}

\section{Conclusion}

We presented a  parallelization scheme 
for the  EC
algorithm and performed extensive tests for 
correctness and efficiency for the 
  hard sphere system in two dimensions.

For parallelization we use a spatial partitioning approach into 
simulation cells. 
We find that it is crucial for correctness  of the PEC algorithm
that the starting particles for ECs in each sweep are drawn {\em with}
 replacement.
We have shown implementations without replacement to give incorrect 
results for the pair correlation function and the resulting 
pressure, see Figs.\ \ref{fig:paar_corr}.
The reason for this incorrect results 
is the violation of detailed balance
on the sweep level.

We analyzed the performance gains for the PEC algorithm
and find the  criterion \eqref{eq:criterion}
for an optimal degree of parallelization. 
Because of the cluster nature of EC moves massive 
parallelization will not be optimal. 
The PEC algorithm will therefore be best suited for 
commonly available multicore CPUs with shared memory.

Finally, we discussed a first application of the algorithm to 
dense polymer systems. Using ECs we simulated 
bundle formation in a  solution of 
attractive semiflexible polymers. 
The EC moves give rise to a faster and much more realistic 
bundle diffusion behavior. This allows us to simulate 
large systems, in particular, if the EC algorithm is parallelized. 
The simulation exhibits large-scale network bundle structures, see Fig.\ 
\ref{fig:network_snapshot}, which are very similar to structures
observed in recent experiments \cite{huber2012,deshpande2012}.

\section*{Acknowledgements}

We acknowledge financial support by  the Deutsche Forschungsgemeinschaft
(KI 662/2-1).

\def\bibsection{\section*{References}}
\bibliography{bib}

\end{document}